# Diagnosis of Paratuberculosis in Histopathological Images Based on Explainable Artificial Intelligence and Deep Learning

Tuncay Yiğit[1], Nilgün Şengöz[1*], Özlem Özmen[2], Jude Hemanth[3], Ali Hakan Işık[4]

[1] Dept. of Computer Engineering, Faculty of Engineering, Süleyman Demirel University, Isparta 32260, Turkey
[2] Dept. of Pathology, Faculty of Veterinary Medicine, Burdur Mehmet Akif Ersoy University, Burdur 15030, Turkey
[3] Karunya Institute of Technology & Sciences, Coimbatore 641114, India
[4] Dept. of Computer Engineering, Faculty of Engineering and Architecture, Burdur Mehmet Akif Ersoy University, Burdur 15030, Turkey

Corresponding Author Email: nilgunsengoz@mehmetakif.edu.tr

**ABSTRACT**

Artificial intelligence holds great promise in medical imaging, especially histopathological imaging. However, artificial intelligence algorithms cannot fully explain the thought processes during decision-making. This situation has brought the problem of explainability, i.e., the black box problem, of artificial intelligence applications to the agenda: an algorithm simply responds without stating the reasons for the given images. To overcome the problem and improve the explainability, explainable artificial intelligence (XAI) has come to the fore, and piqued the interest of many researchers. Against this backdrop, this study examines a new and original dataset using the deep learning algorithm, and visualizes the output with gradient-weighted class activation mapping (Grad-CAM), one of the XAI applications. Afterwards, a detailed questionnaire survey was conducted with the pathologists on these images. Both the decision-making processes and the explanations were verified, and the accuracy of the output was tested. The research results greatly help pathologists in the diagnosis of paratuberculosis.

## 1. INTRODUCTION

Paratuberculosis, a chronic infectious disease of ruminants, is caused by mycobacterium paratuberculosis, and featured by granulomatous inflammation of the intestines and regional lymph nodes. There are some problems in definitive diagnosis of paratuberculosis in living animals. The disease can be difficult to detect, especially by inexperienced and young researchers, for the lack of specific findings in the early stages. The key difficulty lies in the absence of accurate and sensitive diagnostic tests to identify all infected animals. Most of the serological and immunological tests available yield many false positives and negatives, and thus have limited value in diagnostic use [1]. What is worse, there is not yet a vaccine or effective treatment method for paratuberculosis. Veterinary pathology, which deals with the histopathological analysis of animal tissues, plays a key role in diagnostic studies on paratuberculosis.

In this study, the original dataset on paratuberculosis was created Head of Pathology Department, Department of Preclinical Sciences, Faculty of Veterinary Medicine, Burdur Mehmet Akif Ersoy University. The permission from the ethics committee is not required, because the pathological examination was performed on dead animals.

Artificial intelligence and its component machine learning applications are among many latest developments in science and technology. The outstanding performance of artificial intelligence is demonstrated vividly by the rapid proliferation of multi-layer neural network algorithms called deep neural networks (DNNs). Nevertheless, the black box problem emerging in DNNs remains a mystery to those interested in and using artificial intelligence. For key medical imaging systems, artificial intelligence must be more understandable, explainable, and user-oriented, achieving the highest possible human-machine interaction.

A defining feature of artificial intelligence is the ability to reach the context and infer the meaning of a subject or concept from a small dataset. But it is not easy to make sense of a huge amount of fast-flowing, unstructured data. Fortunately, artificial learning and other tools make it possible for artificial intelligence systems to reach suggestions, decisions, or actions from a large amount of data. Yet these tools face many problems, such as the requirement of large high-quality and easily accessible datasets with labels, the need for considerable engineering effort [2], the nonlinear nature of DNNs, and the difficulty in associating data with algorithms.

While many scholars strive to achieve as decision support and autonomous movement with artificial intelligence, it is the subject of explainable artificial intelligence (XAI) that brings a technological definition to explainability for stakeholders like developers, users, decision makers, and controllers. XAI explores and develops the internal logic processes for artificial intelligence systems, and complements artificial intelligence models, making algorithm outputs more explainable to stakeholders.

Medical imaging systems play an important role in the accurate detection and diagnosis of diseases. The physicians or pathologists need to achieve a moment-to-moment all-encompassing state of trust cycle. A medical imaging system enables pathologists to examine the work process from multiple points, handle trust-based relationships actively, and perceive active trust and process evaluation separately.

Pathological imaging differs significantly from other imaging techniques, namely, computed tomography (CT), and magnetic resonance imaging (MRI). Gigapixel images are more complex than other images. It is well-known that large-scale histological image features are quite different from ImageNet photos and radiological images. In this context, this study aims to provide much needed evidence and clarify future research directions concerning the application of deep learning algorithms on histopathological images, and the provision of an explainable level of information to pathologists.

In their daily work, human pathologists must examine and diagnose large numbers of slides from various sources. This is a time-consuming and sometimes impossible task. The same happens to veterinary pathologists. The difficulty of veterinary pathologists in diagnosis can be partly attributed to the fact that very few specialists are experienced in diagnosing the diseases of animals with very different physiological and structural feature. Recently, the diagnostic speed and accuracy of both human and veterinary pathologists are significantly improved by the use of artificial intelligence in pathology.

To investigate the working adequacy and reliability of XAI, expert pathologists must implement a careful and unbiased evaluation to correctly understand the claims outputted by the algorithm. In some cases, the evaluation effect is limited by the time pressure or the insufficient knowledge of the pathologists. This situation may be eliminated with the development of artificial intelligence systems.

The rest of this paper is structured as follows: Section 2 introduces deep learning algorithms and gradient-weighted class activation mapping (Grad-CAM), and our new and original dataset; Section 3 explains the methodology; Section 4 collects the responses from the pathologist participants through a Likert-based questionnaire survey on XAI; Section 5 discusses the survey data; Section 6 provides the conclusions.

## 2. DEEP LEARNING ALGORITHMS AND GRAD-CAM

Based on the standard architectures of artificial neural network (ANN), deep learning techniques aim to solve different problems, and provide various alternative solutions. Typically, a deep learning technique consists of specialized layers for data processing and weighting, and greatly facilitates and even eliminates human-oriented data processing [3-5]. In this way, all kinds of data can be used effectively in artificial intelligence models. For example, image processing, an active field in its own right, has been fully integrated with deep learning and artificial intelligence [6-8].

Medicine is an important application area, as well as the biggest challenge to artificial intelligence and deep learning. The difficulties of applying artificial intelligence and deep learning to medicine include the missing, noisy, inaccurate, and erroneous datasets of high-dimensional images in the medical decision support mechanism [9]. For this reason, researchers are unfortunately deprived of large datasets [10]. To solve the problem, the ultimate solution is to consider the various data on XAI in the context of medicine. Pathologists must understand how and why an artificial intelligence algorithm makes a decision, especially on histopathological images [11].

The research on the application of XAI systems in medicine intends to achieve high learning performance, with the aid of machine learning, and human-machine interaction. There is a conflict between two common metrics of deep learning performance: prediction accuracy and explainability. The best performing deep learning algorithms are the least transparent tools. Meanwhile, the most transparent algorithms, e.g., decision trees, show the least performance [12].

In the last few years, convolutional neural networks (CNNs), first introduced by Fukushima [13], have gained special status as a particularly promising form of deep learning. Rooted in image processing, convolutional layers have successfully penetrated almost all subfields of deep learning [14]. As shown in Figure 1, a typical CNN encompasses convolutional layer(s), pooling or subsampling layer(s), activation function(s), connections, and finally a regulation layer. The convolutional operation mainly attempts to extract feature maps of the input images [15]. CNNs are adopted more commonly than most other networks, especially in image processing applications [16]. This study uses VGGNet and ResNet, which are the most popular CNNs.

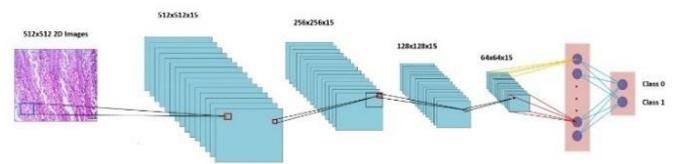

**Figure 1.** CNN architecture

### 2.1 ResNet

ResNet is the CNN that won the Large Scale Visual Recognition Challenge (ILSVRC) of 2015. This architecture adopts special jump connections and batch normalization intensively to reduce parameters and increase the depth, which significantly increases the training success. As shown in Figure 2, ResNet has unique residual modules [17].

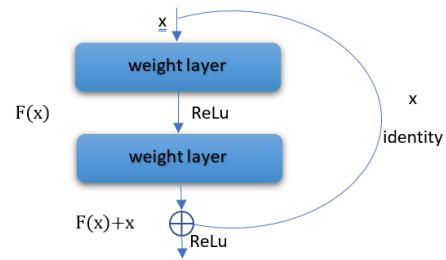

**Figure 2.** ResNet architecture

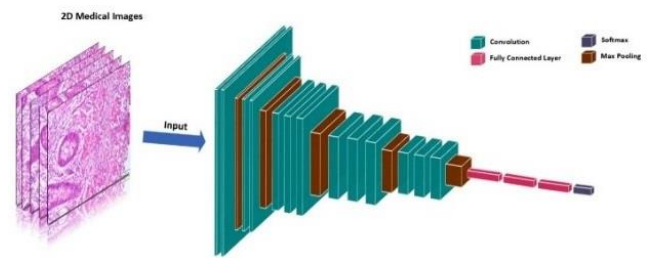

**Figure 3.** The basic building block of the VGG16 network

### 2.2 VGGNet

VGGNet, proposed by Simonyan and Zisserman [18], came second in the 2014 ILSVRC competition. The good performance of the network comes from the critical component: the depth of the mesh. In the entire VGGNet, the network is simplified by maintaining only the convolution process with a 3×3 window and a 2×2 filter, without changing the depth. As shown in Figure 3, the last fully connected layer of VGGNet has 1,000 nodes, and the classification (output) layer has a soft-connected (softmax) layer.

## 2.3 Grad-CAM

Grad-CAM is an important labeling tool that can be applied post-hoc, without any intervention in the deep learning algorithm, i.e., the CNN. The application does not need to change the analyzed network. The tool emerges as a generalization of class activation. The main difference between Grad-CAM and CAM lies in the global average pooling (GAP) and softmax activation after convolutional layers [19]. Grad-CAM provides significant localization for all classes of images. The localizations in an input image are produced by following the gradients in the backpropagation of CNNs. In this way, heatmaps are created on the most prominent features representing a class.

## 3. METHODOLOGY

The original dataset was prepared as follows: Our team conducted extensive research with pathologists in Department of Pathology, Burdur Mehmet Akif Ersoy University to understand the importance of paratuberculosis. During the necropsy, gut samples were collected from 20 control sheep and 20 sheep with natural paratuberculosis, and fixed in 10% buffered formalin. The sections were stained with hematoxylin-eosin, mounted with a coverslip, and examined under a light microscope. The diagnosis was confirmed by Zielh-Neelsen staining for mycobacteria. No pathological findings were observed in the control sheep gut. A marked increase in the thickness of the gut wall was detected in paratuberculosis cases. In addition, a characteristic chronic granulomatous reaction was noticed in propria mucosa. Inflammatory cells were mainly composed of lymphocytes, macrophages, epithelioid cells, and Langhans giant cells. Only Zielh-Neelsen positive cases were included in the paratuberculosis cases.

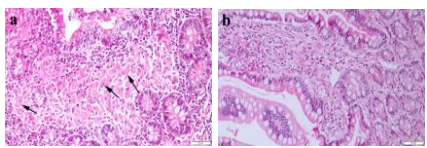

**Figure 4. (a)** Gut from a sheep with paratuberculosis, numerous macrophages (arrows) in submucosae **(b)** healthy intestine, hematoxylin-eosin, scale bars=50µm

Despite the prevalence of paratuberculosis around the world, the changes in the organism regarding the source and development of the disease are not known exactly. Some pathological changes have been recognized as characteristic of paratuberculosis. But they are not sufficient for accurate detection of the disease. The commonly adopted manual histopathological analysis is not only laborious and time-consuming, but also limited by the quality of the specimen and the experience of the pathologists. Therefore, this study takes an important step to eliminate the different decisions both in terms of time and between pathologists in the computer environment. Our dataset consists of 520 microscopic images in two classes, diseased and healthy (Figure 4).

The classification efficiency on the test images was measured by accuracy, sensitivity, and specificity. The metrics were estimated for each class, and the mean value on the two classes was taken to evaluate the performance of the classifier [20]. Accuracy, sensitivity, and specificity can be respectively calculated by:

$$Classification\ Accuracy = \frac{TP + TN}{TP + TN + FP + FN} \quad (1)$$

$$Sensitivity\ (Recall) = \frac{TP}{TP + FN} \quad (2)$$

$$Specificity = \frac{TN}{TN + FP} \quad (3)$$

$$Precision = \frac{TP}{TP + FP} \quad (4)$$

$$F1\ Score = \frac{2 * Precision * Recall}{Precision + Recall} \quad (5)$$

where, TP, TN, FP, and FN are true positive, true negative, false positive, and false negative, respectively. TP and TN return the number from which the classes are predicted correctly, while FP and FN return the number of incorrect predictions of classes with each other.

The accuracy generally reveals the overall relationship, and is often adopted to evaluate the outputs of deep learning algorithms. However, this metric does not appear to be sufficient on its own. This calls for an alternative metric not affected by the problem of unbalanced datasets. Thus, the binary classification performance was also evaluated by Matthews correlation coefficient (MCC), a contingency matrix method for calculating the Pearson product moment correlation coefficient between actual and estimated values. The MCC value can be calculated by:

$$MCC = \frac{TP * TN - FP - FN}{\sqrt{(TP + FP) * (TP + FN) * (TN + FP) * (TN + FN)}} \quad (6)$$

After applying the values obtained from the confusion matrix to the MCC formulation, the worst value was -1, and the best value was +1.

As can be seen from the results in Tables 1-3, our VGG-16 achieved the highest classification accuracy and MCC among three types of CNNs. Figure 5 details the architecture of our VGG-16. After finding the best algorithm, i.e., VGG-16, a heat map, that is, a class activation map, was produced using Grad-CAM.

To the images more transparent and explainable, the trained weights from the proposed VGG-16 were incorporated into the Grad-CAM, an algorithm focusing on how the computer sees, i.e., how images are classified after training in deep learning algorithms. Grad-CAM uses the feature maps produced by the final convolutional layer of a CNN algorithm. The final convolutional layers are expected to have the best explanation between high-level semantics and detailed spatial information. In the test images, which were successfully classified with an accuracy of 0.98, the regions of interest (ROIs) of the VGG16 algorithm were visualized by Grad-CAM (Figures 6 and 7).

**Table 1.** Performance of CNN model

| Classification Accuracy | Sensitivity | Specifity | Precision | F1 Score | MCC | Training Time (s) |
|---|---|---|---|---|---|---|
| 0.96 | 0.97 | 0.95 | 0.95 | 0.96 | 0.91 | 720 |

**Table 2.** Performance of ResNet50

| Classification Accuracy | Sensitivity | Specifity | Precision | F1 Score | MCC | Training Time (s) |
|---|---|---|---|---|---|---|
| 0.97 | 0.97 | 0.98 | 0.98 | 0.97 | 0.95 | 1500 |

**Table 3.** Performance of our VGG-16

| Classification Accuracy | Sensitivity | Specifity | Precision | F1 Score | MCC | Training Time (s) |
|---|---|---|---|---|---|---|
| 0.98 | 0.98 | 0.99 | 0.99 | 0.98 | 0.97 | 585 |

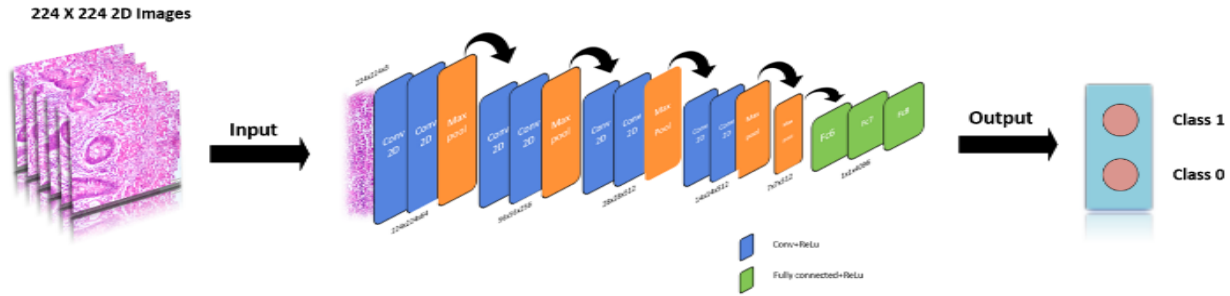

**Figure 5.** Architecture of our VGG-16

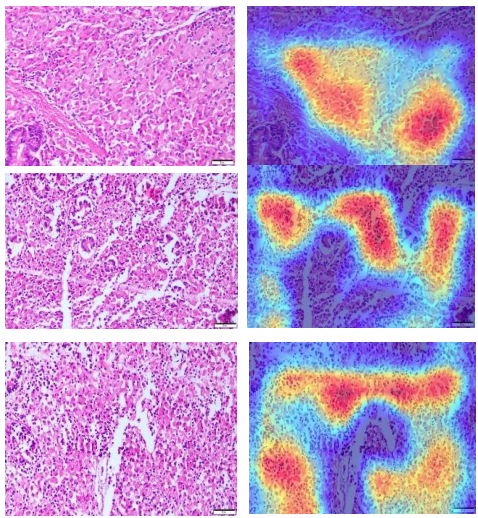 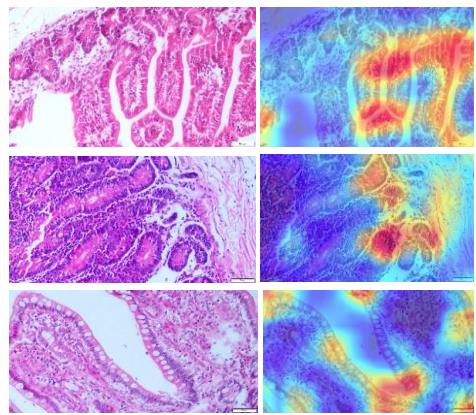

**Figure 6.** Outputs of Grad-CAM on diseased images    **Figure 7.** The outputs of the Grad-CAM on healthy images

## 4. XAI SURVEY

Two scales were used in our survey: the Cahour-Forzy (CHF) scale, and the XAI Explanation Satisfaction scale. The former asks users directly whether they trust the XAI system, and whether the XAI system is predictable, reliable, and efficient [21]. The latter asks users directly whether they trust the XAI system, whether the XAI system is predictable, reliable, efficient, and credible [22]. Tables 4 and 5 list the questions in the CHF scale and XAI Explanation Satisfaction scale, respectively. These questions were asked to expert pathologists.

Before the survey, expert pathologists were first asked about their age, years of experience in the profession, whether they had done any work on artificial intelligence before, and whether they were involved in such a study. As shown in Table 6, the mean age was 32.4, the mean year of experience was 10.3, and only 2 had any experience in artificial intelligence.

In the next stage of the survey, a presentation titled "What is Artificial Intelligence and How Computers See" was given to the pathologists to resolve any doubt. Afterwards, four visual presentations of the paratuberculosis disease within the scope of the study were presented, and the pathologists were asked to predict the disease. Since the image of the disease was reflected in the presentation, time was kept with a stopwatch. After a total of 3min44s, every pathologist wrote the disease in the survey study. In the end, only 3 pathologists predicted the disease correctly, while 7 predicted the disease incorrectly or left it blank.

**Table 4.** Results for CHF scale of the expert survey evaluation

| No. | Statement | Average feedback |
|---|---|---|
| CHF1 | What is your confidence in the artificial intelligence system? Do you have a sense of trust in the system? | 3.9 |
| CHF2 | Are the actions of the artificial intelligence system predictable? | 4.5 |
| CHF3 | Are the artificial intelligence system reliable? Do you think it is safe? | 4 |
| CHF4 | Do you think the artificial intelligence system is efficient in what it does? | 5.1 |

Results are the means of a total of 10 participating veterinary doctors.

**Table 5.** Results for XAI scale of the expert survey evaluation

| No. | Statement | Average Feedback |
|---|---|---|
| XAI1 | I trust the artificial intelligence system. I feel like it's working well. | 3.6 |
| XAI2 | The outputs of the artificial intelligence system are very predictable. | 3.4 |
| XAI3 | The artificial intelligence system is very reliable. I can always count on it to be true. | 2.8 |
| XAI4 | When I trust the artificial intelligence system, I feel confident and good that I will get the right answers. | 3 |
| XAI5 | The artificial intelligence algorithm is efficient, because it works very fast. | 4 |
| XAI6 | I am cautious about artificial intelligence. | 4.1 |
| XAI7 | Artificial intelligence can perform histopathological image analysis better than a novice human user. | 4.1 |
| XAI8 | I prefer to use the artificial intelligence system to make decisions. | 3.9 |
| XAI9 | Artificial intelligence covers all relevant information to help me to decide. | 2.5 |
| XAI10 | Artificial intelligence helps me make faster decisions. | 3.4 |
| XAI11 | EAI helps me to decide. | 3.5 |
| XAI12 | The overall detection and uptime performance of the system is good. | 3.8 |
| XAI13 | This model can be embedded in a broader healthcare support system such as the Internet of things (IoT). | 4.2 |
| XAI14 | This system can be used in further treatment and diagnostic strategies. | 4 |
| XAI15 | This model is not good at diagnosing histopathological diseases. | 2.8 |

Results are the means of a total of 10 participating veterinary doctors.

**Table 6.** Summary for expert survey evaluation

| Age | Work experience (years) | AI experience | | Prediction | |
|---|---|---|---|---|---|
| | | Yes | No | True | False/No pred. |
| 34.2 | 10.3 | 20% | 80% | 30% | 70% |

## 5. RESULTS AND DISCUSSION

The reliability and normal distribution of the two scales were calculated. As shown in Table 7, the mean of the CHF scale, and XAI Explanation Satisfaction scale was 4.33 and 3.51, respectively. This means the pathologists find the outputs of the Grad-CAM application sufficient and useful in the future.

**Table 7.** Reliability and correlation results

| | Mean | S.D. | 1 | 2 | Skewness | Kurtosis |
|---|---|---|---|---|---|---|
| 1.CHF | 4,33 | ,94 | **.63** | | ,700 | -,288 |
| 2.XAI | 3,51 | ,52 | ,748* | **.75** | ,412 | -,110 |

*: $p < .05$
Cronbach's Alpha reliability values are shown in bold.
S.D.: Standard Deviation

Skewness and kurtosis tests were conducted to see if a distribution is normal. Note that skewness measures the degree of asymmetry of a distribution, and kurtosis indicates the degree to which too many or too few samples appear in the middle of the distribution. According to the skewness and kurtosis results in Table 7, our dataset belongs to the normal distribution, for the two metrics fell between -1 and 1.

The Cronbach's alpha (α), which measures the internal consistency of the items, was used to verify the homogeneity of the items in each scale. A high Cronbach's α means the items in a scale are consistent with each other and measure the same feature. The results show that the CHF scale (α=0.63) and XAI Explanation Satisfaction scale (α=0.75) values were at acceptable levels. In additions, a significant positive correlation was found between the two scales (p<.05).

**Table 8.** Regression results

| | (n=10) XAI | | | |
|---|---|---|---|---|
| | β | S.D | p | Summary of Regression Model |
| CHF | .75 | .13 | .02 | $R^2$ =.56, Adj.$R^2$ =.50, $p$=.02, F=8.88 |

Table 8 shows the regression results. It can be observed that the independent variable of the CHF scale has a significant positive effect on the dependent variable of XAI Explanation Satisfaction scale (β=0.75, S.E.=0.13). 50% of the variance in XAI Explanation Satisfaction scale is explained by the independent variable of the CHF scale (Adjusted $R^2$ =0.50).

As shown in Table 9, a significant difference was found in the mean of the CHF scale in terms of professional experience difference (t=-3.315, p<0.05). Accordingly, the mean of the participants with 5 years or more of professional experience (X̄=5.17, S.D.= 0.88) was found to be significantly higher than the mean of those with 4 years or less (X̄=3.78, S.D.=0.46). No significant difference was found in the mean of the variable of XAI Explanation Satisfaction scale (p>0.05).

The fact that our CNN models performed at least 95% on the three metrics, especially against the new and original dataset on paratuberculosis disease, is remarkable in terms of the applicability. But this finding does not directly express the level of safety and comprehensibility of the system. Then, it was deemed necessary to interpret the findings in the evaluations in the context of artificial intelligence and human-machine interaction that can be explained by 10 different pathologists.

**Table 9.** Difference test results

| | 4&Under | | 5+ | | | | |
|---|---|---|---|---|---|---|---|
| | $\bar{X}$ | S.D. | $\bar{X}$ | S.D. | p | t | Significant Difference |
| CHF | 3.78 | .46 | 5.17 | .88 | .01 | -3.31 | 5+>4&Under |
| XAI | 3.27 | .46 | 3.82 | .46 | .15 | -1.79 | None |

Therefore, Grad-CAM was applied to the outputs of the VGG16 model, which has the best classification accuracy. The heat maps thus created were interpreted by the pathologists as highly accurate. Positive feedbacks were given in terms of performance and XAI. Putting aside the detection performance of paratuberculosis disease, this system can be explained. In other words, XAI system was proved as an effective tool in the field of health, thanks to its safety for humans and effectiveness.

In addition, the alternative parameter testing (even hyperparameter optimization) could be evaluated on the VGG16 model within the scope of open problem applications. Although the images selected in the study are preferred for a balanced solution, applications can be made with alternative datasets composed of all images in different studies or the same number of different images. Alternative images can be used apart from the selected images (this is also within the scope of the current research, and acceptable in the context of the open problem). Apart from histological images, the established CNN-VGG16/Grad-CAM system is applicable to other medical images.

Today's technologies are also advancing towards the IoT, and, in this context, the Internet of Health Things (IoHT). Our approach provides an interface for a typical IoHT solution, and create an effective hardware-software synergy, thanks to the explainability factor that user needs.

## 6. CONCLUSIONS

At present, deep learning algorithms are widely and successfully utilized in many fields, such as autonomous driving, face recognition, and chatbots. But serious doubts have been casted on how the current algorithms make decisions, i.e., the black box problem. Even if the mathematical expressions and theories are solved for artificial intelligence-based algorithms, there are serious problems in the explainability and interpretability of such models. Admittedly, models like decision trees have a high level of explainability. Yet they cannot compete with deep learning algorithms in terms of classification accuracy.

This study creates heat maps of the VGG16 model, which is most accurate classifier among three CNNs, with Grad-CAM. On this basis, a serious research proposal was presented to pathologists about where computers actually focus during classification, and whether these focal points are correct. According to the survey results on ten pathologists, the artificial intelligence algorithm focused on the correct places, namely, focusing on the same people during the classification. As a result, the algorithm succeeded in producing a highly successful solution. This reflects the promising performance of the VGG16 model. With the help of the developed system heat maps, it is safe to say that we provided a competent and explainable diagnostic tool for the detection of paratuberculosis disease.

In future, it is crucial to facilitate human-machine interaction smoothly and effectively in artificial intelligence-based systems, especially in medical imaging systems. Besides, it is indispensable to realize an explainable interaction based on trust. While providing the necessary explanation to users, future artificial intelligence-based systems should imitate how the users produce and convey explanations to any situation encountered in their daily life. This study lays the basis for future research on alternative diagnostic processes, and even contributes to the literature with different evaluations using different explainable artificial intelligence techniques.

## ACKNOWLEDGMENT

We would like to thank the pathologists at Burdur Mehmet Akif Ersoy University who contributed to this study with their valuable opinions.

## REFERENCES

[1] Chiodini, R.J., Van Kruiningen, H.J., Merkal, R.S. (1984). Ruminant paratuberculosis (Johne's disease): The current status and future prospects. Cornell Vet., 74(3): 218-262.

[2] Holzinger, A. (2018). From machine learning to explainable AI. 2018 World Symposium on Digital Intelligence for Systems and Machines (DISA), pp. 55-66. https://doi.org/10.1109/DISA.2018.8490530

[3] Şengöz, N., Yiğit, T., Özmen, Ö., Isık, A.H. (2022). Importance of preprocessing in histopathology image classification using deep convolutional neural network. Advances in Artificial Intelligence Research, 2(1): 1-6. https://doi.org/10.54569/aair.1016544

[4] Dara, S., Tumma, P. (2018). Feature extraction by using deep learning: A survey. International Conference on Electronics, Communication and Aerospace Technology (ICECA), Coimbatore: India, pp. 1795-1801. https://doi.org/10.1109/ICECA.2018.8474912

[5] Du, X., Cai, Y., Wang, S., Zhang, L. (2016). Overview of deep learning. Youth Academic Annual Conference of Chinese Association of Automation (YAC), Wuhan: China, pp. 159-164. https://doi.org/10.1109/YAC.2016.7804882

[6] Jiao, L., Zhao, J. (2019). A survey on the new generation of deep learning in image processing. IEEE Access, 7: 172231-172263. https://doi.org/10.1109/ACCESS.2019.2956508

[7] Çınar, A., Yıldırım, M., Eroğlu, Y. (2021). Classification of pneumonia cell images using improved ResNet50 model. Traitement du Signal, 38(1): 165-173. https://doi.org/10.18280/ts.380117

[8] Hassaballah, M., Awad, A.I. (2020). Deep Learning in Computer Vision: Principles and Applications. Vol. 1, CRC Press, Boca Raton: USA. https://doi.org/10.1201/9781351003827

[9] Holzinger, A., Dehmer, M., Jurisica, I. (2014).

Knowledge discovery and interactive data mining in bioinformatics—State-of-the-art, future challenges and research directions. BMC Bioinformatics, 15: I1. https://doi.org/10.1186/1471-2105-15-S6-I1

[10] Holzinger, A. (2016). Interactive machine learning for health informatics: When do we need the human-in-the-loop? Brain Informatics, 3(2): 119-131. https://doi.org/10.1007/s40708-016-0042-6

[11] Holzinger, A., Biemann, C., Pattichis, C.S., Kell, D.B. (2017). What do we need to build explainable AI systems for the medical domain? arXiv:1712.09923.

[12] Bologna, G., Hayashi, Y. (2017). Characterization of symbolic rules embedded in deep DIMLP networks: A challenge to transparency of deep learning. Journal of Artificial Intelligence and Soft Computing Research, 7(4): 265-286. https://doi.org/10.1515/jaiscr-2017-0019

[13] Fukushima, K. (1980). Neocognitron: A self-organizing neural network model for a mechanism. Biol. Cybernetics, 36(4): 193-202. https://doi.org/10.1007/BF00344251

[14] LeCun, Y., Bottou, L., Bengio, Y., Haffner, P. (1998). Gradient-based learning applied to document recognition. Proceedings of the IEEE, 86(11): 2278-2324. https://doi.org/10.1109/5.726791

[15] Lecun, Y., Bengio, Y., Hinton, G. (2015). Deep learning. Nature, 521(7553): 436-444. https://doi.org/10.1038/nature14539

[16] Coşkun, M., Yıldırım, Ö., Uçar, A., Demir, Y. (2017). An overview of popular deep learning methods. European Journal of Technic (EJT), 7(2): 165-176. https://doi.org/10.23884/ejt.2017.7.2.11

[17] He, K., Zhang, X., Ren, S., Sun, J. (2016). Deep residual learning for image recognition. 2016 IEEE Conference on Computer Vision and Pattern Recognition (CVPR), pp. 770-778. https://doi.org/10.1109/CVPR.2016.90

[18] Simonyan, K., Zisserman, A., (2014). Very deep convolutional networks for large-scale image recognition. arXiv:1409.1556.

[19] Selvaraju R.R., Cogswell M., Das, A., Vedantam, R., Parikh, D., Batra, D. (2017). Grad-CAM: Visual explanations from deep networks via gradient-based localization. IEEE International Conference on Computer Vision (ICCV), pp. 618-626. https://doi.org/10.1109/ICCV.2017.74

[20] Hemanth, D.J., Vijila, C.K.S., Selvakumar, A.I., Anitha, J. (2013). Performance improved iteration-free artificial neural networks for abnormal magnetic resonance brain image classification, Neurocomputing, 130: 98-107 https://doi.org/10.1016/j.neucom.2011.12.066

[21] Cahour, B., Forzy, J.F. (2009). Does projection into use improve trust and exploration? An example with a cruise control system. Safety Science, 47(9): 1260-1270. https://doi.org/10.1016/j.ssci.2009.03.015

[22] Hoffman, R.R., Mueller, S.T., Klein, G., Litman, J. (2018). Measuring trust in the XAI context. Technical Report, DARPA Explainable AI Program.